\documentstyle[12pt]{article}
\newcommand{\be}{\begin{equation}}
\newcommand{\ee}{\end{equation}}
\newcommand{\rref}[1]{(\ref{#1})}
\begin{document}

\begin{flushright}
ULB--TH--97/19 \\
hep-th/9710027 \\
October 1997\\
\end{flushright}

\vspace{.8cm}

\begin{center}
{\huge Little Theories in Six and Seven}

\vspace{.4cm}
{\huge Dimensions}
\vspace{1.5cm}

{\large Riccardo Argurio}\footnote{
Aspirant F.N.R.S. (Belgium). E-mail:
rargurio@ulb.ac.be}
{\large and Laurent Houart}\footnote{
Chercheur I.I.S.N. (Belgium). E-mail: lhouart@ulb.ac.be}
\addtocounter{footnote}{-2}\\
\vspace{.4cm}
{\it Service de Physique Th\'eorique}\\
{\it Universit\'e Libre de Bruxelles, Campus Plaine, C.P.225}\\
{\it Boulevard du Triomphe, B-1050 Bruxelles, Belgium}\\

\end{center}

\vspace{1.5cm}

\begin{abstract}

We discuss theories with 16 and 8 supercharges in 6 and 7 dimensions.
These theories are defined as world-volume theories of 5- and 6-branes
of type II and M theories, in the limit in which bulk modes decouple.
We analyze in detail the spectrum of BPS extended objects of these
theories, and show that the 6 dimensional ones can be interpreted as
little (non-critical) string theories. The little 5-branes of the 6 
dimensional theories with 16 supercharges are used to find new string
theories with 8 supercharges, which have additional group structure. 
We describe the web of dualities relating all these theories.
We show that the theories with 16 supercharges
can be used for a Matrix description of M-theory on $T^6$ in the general
case, and that they also reproduce Matrix theory on $T^5$ and $T^4$ in
some particular limit.

\end{abstract}

\newpage

\section{Introduction}

A promising approach to the understanding of M-theory \cite{huto,witt,demo}
is the so-called M(atrix) theory \cite{bfss}. According to this original
proposal, the supersymmetric $U(N)$ matrix quantum mechanics of $N$ D0-branes
describes M-theory in flat 11 dimensional space, in the infinite momentum
frame (IMF) when $N \rightarrow \infty$ or alternatively in the discrete light
cone quantization \cite{suss} for finite $N$. M-theory toroidal 
compactifications
are described by an equivalent M(atrix) model in which the matrix quantum
mechanics is replaced by super Yang-Mills (SYM) \cite{bfss,tay1,snd,tay2} on
a dual torus. However when there are more than three compact dimensions
the SYM is ill-defined because it is non-renormalizable (see e.g.\cite{sei1}).
In order to circumvent this problem one has to go beyond the SYM prescription.
Matrix theory on $T^4$ is described in terms of a $(2,0)$ field theory in 5+1 
dimensions \cite{roza1,roza2}, which corresponds to the theory on the 
world-volume of $N$ coinciding M5-branes \cite{stro}. 
Compactifying further one has 
to abandon the idea of having a field theory description. On $T^5$, 
Matrix theory is
believed to be described in terms of a non-critical string theory in 5+1 
dimensions 
\cite{roza2,sei2} obtained from $N$ NS 5-branes at vanishing type II string 
coupling.
On $T^6$, a description using the M-theory KK 6 monopole has been 
recently proposed \cite{brun,hana}. It appears that this
6+1 dimensional theory contains membranes, and it has been called 
``m-theory"
\cite{mms}. 

The similarity between m-theory and M-theory is actually striking. One can 
indeed
define ``little string theories" in 5+1 dimensions, one chiral and one 
non-chiral, and relate them to m-theory by T-dualities and 
(de)compactification.
These theories can be defined using 5-branes of several types appearing in
type II and M theories\footnote{
This approach was pioneered in the work of \cite{dijk}.}, 
always with additional transverse compact directions
\cite{mms}. It has to be stressed that these additional compact directions
introduce new parameters with respect to the theories defined by Seiberg 
\cite{sei2},
thus making them suitable for a description of Matrix theory on $T^6$.

In this paper, we find interesting to study the little string theories and 
m-theory
in their own respect. We first revisit the theories in 6 and 7 dimensions with
16 supercharges leading to $ii$a, $ii$b little string theories and m-theory. 
The different ways to obtain these theories are analyzed. We start 
from 5 and 6 
dimensional extended objects defined in M or type II 
theories and we take limits in which bulk modes decouple. 
This leads nevertheless to a non-trivial theory without gravity
defined on the world-volume of the extended objects. We show the web of 
dualities
between these little theories which exactly reproduces the scheme of 
the ``big" 
theories in 10 and 11 dimensions. In a Matrix theory perspective, 
the spectrum of the BPS extended
objects of these little theories is investigated and it is shown that it 
agrees 
with the U-duality group of M-theory compactified on $T^6$. 
Furthermore the theories used to describe M-theory on $T^4$ and $T^5$ are 
recovered 
as particular limits of these little string theories.

We then turn to theories in 6 dimensions with 8 supercharges. These theories 
have
(1,0) supersymmetry, do not contain gravity and may have an additional gauge
symmetry. Our strategy is to obtain them from the theories with 16 
supercharges.
We mimic the 10 dimensional procedure in which type I theory
is obtained from IIB theory introducing an $\Omega 9$ orientifold and 16
D9-branes
\cite{sagn,petr,polc} (see also \cite{hull}). 
The two heterotic string theories are then
found by chains of dualities. Applying the same procedure to the 6 dimensional
theories, we find one theory with open strings and two with closed strings, 
which
we call respectively type $i$, $h_b$ and $h_a$ theories. These are in fact 
classes
of theories. Unlike the 10 dimensional case, the gauge group is  not 
constrained.
Moreover, there is no simple description of the gauge theory defined by the
$h_a$ ``little heterotic" theories, since this is again related to the (2,0) 
theory.
As a consistency check of the picture, the $h_a$ theory can also be related to
a particular compactification of m-theory.

The paper is organized as follows. In section 2 we study the theories with
16 supercharges. For each one of the three theories we review the different 
ways
to obtain them and explain how the limits taken are related by a chain of 
dualities. 
The limits take a particular simple form in the formulation which uses
the KK monopoles. The relation between the little theories and 
compactifications
of Matrix theory on tori is explained. In section 3, we consider the theories 
with 
8 supercharges. We propose the definition of little type $i$ open string 
theory,
and of two kinds of little ``heterotic" theories. The final section contains
a brief discussion.

\section{Theories with 16 supercharges}

Supersymmetric theories with 16 supercharges naturally appear in type II 
string
theories and M-theory as the effective theory on the world-volume of BPS 
branes. 
In order to have well-defined theories on world-volumes one has to take a 
limit
in which the bulk modes decouple. This is achieved by sending the Planck mass, 
defined with respect to the non-compact space, to infinity. 

We will consider here theories defined on the world-volume of 5 branes
and 6 branes, and such that at least three of the transverse directions are
non-compact (in order to keep the space asymptotically flat). 
This allows for extra transverse compact directions, which will
actually play a key r\^ole in defining the parameters of the little theories.

In M-theory, we have the following two objects:
\begin{itemize}
\item M5-brane, with up to 2 transverse compact directions parametrized by 
$R_1$ and $R_2$.
\item KK6-brane, which has naturally a transverse compact direction, the
so-called NUT direction (see \cite{sen} for a recent review on KK monopoles).
\end{itemize}
In type IIA theory we have the following three objects:
\begin{itemize}
\item NS5(A)-brane, with 1 transverse compact direction parametrized by 
its radius $R_A$.
\item KK5(A)-brane, with its transverse NUT compact direction.
\item D6-brane, with no compact transverse directions.
\end{itemize}
The objects we have in type IIB theory are:
\begin{itemize}
\item NS5(B)-brane, with 1 transverse compact direction.
\item KK5(B)-brane, with its NUT compact direction.
\item D5-brane, with 1 compact transverse direction.
\end{itemize}

All these branes are related by the usual dualities relating type II and 
M-theory.
We will however distinguish between dualities which leave the world-volume
of the branes unaffected, as transverse T-dualities for NS branes 
(NS5 and KK5), 
IIB S-duality and transverse
compactifications, and dualities which on the other hand act on the 
world-volume,
as T-dualities and compactifications along a world-volume direction of 
NS branes and T-dualities for D-branes.

Considering the dualities leaving the world-volume unaffected leads to three
different families of branes each one defining one theory:
\begin{itemize}
\item{$ii$a:} KK5(A) $\leftrightarrow$ NS5(B) $\leftrightarrow$ D5
\item{$ii$b:} KK5(B) $\leftrightarrow$ NS5(A) $\leftrightarrow$ M5
\item{m:} KK6 $\leftrightarrow$ D6
\end{itemize}
These three theories are related by dualities which affect the world-volume 
of the branes. A T-duality along the world-volume of a NS5 or a KK5
changes from IIA to IIB and thus also from $ii$a to $ii$b. Compactification of
the KK6 on one of its world-volume directions yields the KK5(A), thus relating
$ii$a and m theories via compactification. The same duality between little
theories is obtained acting with a T-duality on the world-volume of the D6,
which gives the D5. Note also that the D4-brane, which defines a theory in
5 dimensions, can be obtained either by a T-duality from the D5, or by
compactification from the M5. This shows that once compactified, there is no
longer difference between $ii$a and $ii$b theories in 5 dimensions.

Although the relations discussed above are rather formal at this stage, they
exactly reproduce the same pattern of dualities of the 10 and 11 dimensional
theories. We will show hereafter that in the proper limits in which
the above little theories make sense (i.e. they decouple from the bulk), 
this structure still holds and aquires even more evidence.

We now turn to the description of the different little theories.

\subsection{$ii$a theory}

As explained above, there are three ways to define type $ii$a theory \cite{mms}. 
The six
dimensional supersymmetry is (1,1). This is most easily found for the D5 brane
from dimensional reduction of the $N=1$ supersymmetry in $D=10$ \cite{tasi}.
For the NS5(B) and the KK5(A) it has been discussed respectively in \cite{chs}
and \cite{sen}. The type $ii$a theory is thus non-chiral.

The first approach is based on the D5 with a 
transverse compact direction of radius $R_B$. 

We look for all the objects which from the D5 world-volume point of view have
a finite tension, i.e. we rule out branes extending in transverse non-compact
directions. The relevant configurations of branes intersecting with the D5, 
and breaking further $1/2$ of
the supersymmetry are: D1$\subset $D5, F1$\mapsto$D5, D3$\mapsto$D5, 
NS5(B)$\mapsto$D5 and KK5(B)$\parallel$D5. The F1, D3 and NS5 have a 
boundary on
the D5 \cite{stro,aehw}, and their only dimension 
transverse to it wraps around the transverse compact direction.
 
Generically, supergravity solutions preserving $1/4$ of the supersymmetries
and representing two intersecting branes can
be computed \cite{tsey,kast,vand,aeh}. Their existence can
be deduced by the compatibility of the two supersymmetry projections which 
characterize the configuration. 
The supersymmetry projections characterizing the branes are discussed in the
appendix, where we also fix the notations.

Before taking the limit in which the bulk decouples, we have to fix the 
tension and the coupling of the little string theory on the world-volume
of the D5-brane. Since we have three parameters at hand, namely the string
length $l_s$, the string coupling of IIB theory $g_B$ and the radius $R_B$, it
will be possible to send the 9 dimensional Planck mass\footnote{
We have to consider the Planck mass in 9 dimensions because one of the 
transverse directions is compact. Furthermore its radius will
be sent to zero in the limit discussed above. Note also that this limit
does not depend on the size of the 
directions longitudinal to the D5-brane. For simplicity, we take them 
to be infinite.
} to infinity while
keeping a non-trivial little theory on the brane.

The only string-like object which lives on the D5-brane is the D1 string 
trapped to its world-volume \cite{doug}.
We take it to define the fundamental little string of $ii$a theory.
Accordingly, its tension is defined by (using \rref{tdbrane}):
\be t_a\equiv T_{D1}={1 \over g_B l_s^2}. \label{ltfa} \ee
The boundaries of the F1, D3 and NS5, which are respectively 0-, 2- and 
4-dimensional closed objects, act as little ``d-branes" for the $f1$
little string. Their tension is postulated to be inversely proportional to
coupling of the little string theory $g_a$ \cite{mms}. We have:
\be
\begin{array}{rcccccl}
t_{d0}&\equiv& T_{F1}R_B & = & {R_B \over l_s^2} & \equiv & 
{t_a^{1 / 2} \over g_a} \\
t_{d2}&\equiv& T_{D3}R_B & = & {R_B \over g_B l_s^4} & \equiv & 
{t_a^{3 / 2} \over g_a} \\
t_{d4}&\equiv& T_{NS5}R_B & = & {R_B \over g_B^2 l_s^6} & \equiv & 
{t_a^{5 / 2} \over g_a} 
\end{array}
\label{ltda}
\ee
The above definitions are consistent and, taking into account \rref{ltfa} 
we have:
\be g_a={l_s \over g_B^{1 / 2} R_B}. \label{lga} \ee

The last object to consider is the KK5, which actually fills the world-volume
of the D5. We can nevertheless define its tension using \rref{tkk5}:
\be t_{s5}\equiv T_{KK5}={R_B^2 \over g_B^2 l_s^8}={t_a^3 \over g_a^2 }. 
\label{tsa}
\ee
The $d$4 and $s$5 branes were overlooked in the analysis of \cite{mms},
they are however defined by perfectly well-behaved 10 dimensional
configurations. They are important in the identification of this little
theory as a model for a toroidal compactification of Matrix theory as we
discuss at the end of this section.

We have defined the string tension $t_a$ and the  string coupling $g_a$ of 
the little theory. In order for this $ii$a theory to make sense, we have to
take a limit in which the bulk modes decouple i.e. a limit in which 
the nine dimensional Planck Mass $M_p$ is going to infinity at fixed  $t_a$ 
and 
 $g_a$. The Planck Mass is given by:
\be
M_p^7 = {R_B \over g_B^2 l_s^8}=
{g_B t_a^{7 / 2} \over g_a}.
\label{mplanck}
\ee
The limit defining type $ii$a is thus characterized by:
\be 
g_B \rightarrow \infty , \qquad l_s \rightarrow 0 , \qquad R_B \rightarrow 0
\label{liia}
\ee

We can also find the $ii$a theory starting with the NS5(B)-brane with one
transverse compact direction of radius $\tilde{R}_B$. 
We call, in this case, the string
coupling of type IIB ${\tilde g}_B$ and the string length ${\tilde l}_s$. 
The 10 dimensional configurations breaking 1/4 supersymmetry which define 
the BPS objects living in 6 dimensions
are simply obtained by S-duality from the ones discussed in the preceding
approach. They are the following: F1$\subset $NS5(B), D1$\mapsto$NS5(B), 
D3$\mapsto$NS5(B), D5$\mapsto$NS5(B) and KK5(B)$\parallel$NS5(B). 
The little $ii$a
string is identified to the fundamental string of type IIB theory. Its
tension is simply given by: 
$t_a = \tilde{l}_s^{-2}$. The little string coupling is:
\be g_a= {{\tilde g}_B \tilde{l}_s \over \tilde{R}_B}. \label{gns5b} \ee 
It can be obtained for instance computing $t_{d0}=T_{D1} \tilde{R}_B$.
The limit in which the Planck mass goes to infinity is defined by 
${\tilde g}_B 
\rightarrow 0$, $\tilde{R}_B \rightarrow 0$ and $\tilde{l}_s$ constant. 
This result is 
consistent with the S-duality transformations:
$g_B \rightarrow {\tilde g}_B= 1/ g_B,
 l_s^2 \rightarrow \tilde{l}_s^2= g_B l_s^2$ and $\tilde{R}_B=R_B$ left 
unchanged.

The third object with a 6 dimensional (1,1) supersymmetric world-volume
which can be used to define $ii$a theory is the KK5 monopole of type
IIA string theory, obtained by a T-duality on the transverse compact
direction from the NS5(B)-brane. This direction becomes the NUT direction
of the Euclidean Taub-NUT space transverse to the KK5 world-volume \cite{sen}.
It appears that in this picture all the relevant configurations which
preserve $1/4$ supersymmetries are made up from branes of type IIA
inside the world-volume of the KK5(A) \cite{mms}: FI$\subset$KK5(A),
D0$\subset$KK5(A), D2$\subset$KK5(A), D4$\subset$KK5(A) and
NS5(A)$\subset$KK5(A). This makes the identifiaction of $t_a$ and $g_a$
straightforward. The fundamental $ii$a string coincides now with 
type IIA's F1, and thus $t_a={\tilde l}_s^{-2}$. 
Since here also the ``little"
$d$-branes coincide with the D-branes of type IIA (with $p\leq 4$),
also the little string coupling is given by the IIA one: $g_a=g_A$.
It is easy to find by T-duality from the NS5(B) picture the limit
in which the KK5 decouples from the bulk. Since under T-duality 
$g_B \rightarrow g_A={g_B {\tilde l}_s \over R_B}$, $R_B \rightarrow R_A=
{{\tilde l}_s^2 \over R_B}$ and ${\tilde l}_s$ is unchanged, 
in the KK5(A) picture
we have $g_A$ constant and $R_A \equiv R_{\mbox{NUT}} \rightarrow \infty$.
The Riemann tensor of the Taub-NUT geometry vanishes in this limit,
an indication that the KK monopole decouples from bulk physics.

We recapitulate the BPS spectrum of type $ii$a theory in the following table.
We  list the different little branes and their mass considering now a compact
world-volume characterized by radii $\Sigma_i$ with $i=1 \dots 5$ and volume 
$\tilde{V}_5=\Sigma_1 \dots \Sigma_5$.
We include for later convenience the KK momenta $w$.
\[
\begin{array}{|c|c|} \hline 
\mbox{Brane} & \mbox{Mass} \\ \hline
w & {1 \over \Sigma_i} \\
& \\
f1 & \Sigma_i t_a \\
& \\
d0 & {t_a^{1 / 2} \over g_a} \\
& \\
d2 & {\Sigma_i \Sigma_j t_a^{3 / 2} \over g_a} \\
& \\
d4 & {\tilde{V}_5 t_a^{5 / 2} \over \Sigma_i g_a} \\
& \\
s5 & {\tilde{V}_5 t_a^3 \over g_a^2} \\
& \\ \hline
\end{array} 
\]
\centerline{Table 1: mass of the BPS objects in $ii$a theory.}

We also summarize below the different ways to obtain $ii$a theory
and the relation between the parameters.
\[
\begin{array}{|c|c|c|c|} \hline & & & \\
& D5 & NS5(B) & KK5(A) \\  
& {1 \over g_B}, R_B, l_s \rightarrow 0 & {\tilde g}_B, \tilde{R}_B 
\rightarrow 0 
& R_{NUT} \rightarrow \infty \\ & & & \\ \hline & & & \\
t_a & {1 \over g_B l_s^2} & {1 \over \tilde{l}_s^2} & 
{1 \over {\tilde l}_s^2} \\ & & & \\
g_a & {l_s \over g_B^{1 / 2} R_B} & {{\tilde g}_B \tilde{l}_s \over 
\tilde{R}_B} & g_A \\ & & & \\
\hline \end{array} 
\]
\centerline{Table 2: definitions of $ii$a parameters.}

\subsection{$ii$b theory}

We recall that there are three approaches to this 6 dimensional theory,
using respectively the M5-brane with two transverse compact directions,
the NS5-brane of type IIA with one compact transverse direction and
the KK5 monopole of type IIB \cite{mms}. These three different branes all have 
a world-volume theory with (2,0) chiral supersymmetry \cite{chs,wten,stro,sen}.

The procedure by which we analyze the structure of $ii$b little string
theory is similar to the one described in the preceding subsection.
We will however meet here an interesting structure of $ii$b which is
its $s$-duality. We begin with the M5 approach, where this duality
is geometric.

The M5-brane set up is characterized by the 11 dimensional Planck length
$L_p$ and the two radii $R_1$ and $R_2$ of the two transverse compact
directions. The configurations breaking 1/4 supersymmetry in M-theory 
leading to finite tension
objects on the world-volume of the M5 are the following:
M2$\mapsto$M5 with the M2 direction orthogonal to the M5 wrapping either
$R_1$ or $R_2$; M5$\cap$M5=3; KK6$\supset$M5 with the NUT direction of
the KK6 identified either with $R_1$ or $R_2$. 

The boundaries of the M2-branes are strings on the M5, but we cannot
immediately identify the fundamental $ii$b little string because 
we have two different kinds of them. We simply choose one of the two
(say, the boundary of the
M2 wrapped on $R_1$) to be the fundamental and thus to have
tension $t_b$, and the other to be the little $d$1 brane with tension
${t_b \over g_b}$. This defines $g_b$. $s$-duality of $ii$b is then
simply the interchange in M-theory of $R_1$ and $R_2$ (this can actually
be extended to a full $SL(2,Z)$ duality group considering M2-branes
wrapped on $(p,q)$ cycles of the torus). We have thus (cfr. \rref{tmbranes}):
\be
\begin{array}{rcl}
t_{f1} =& T_{M2} R_1 = {R_1 \over L_p^3} &\equiv t_b, \\
t_{d1} =& T_{M2} R_2 = {R_2 \over L_p^3} &\equiv {t_b \over g_b}.
\end{array} \label{tfb}
\ee
The little string coupling is then given by:
\be g_b={R_1 \over R_2}. \label{lgb} \ee
We can now identify the other world-volume objects by their tension:
\be
\begin{array}{rcl}
T_{M5} R_1 R_2 = &{R_1 R_2 \over L_p^6}= {t_b^2 \over g_b} &\equiv t_{d3} \\
T_{KK6} R_2 =& {R_1^2 R_2 \over L_p^9} = {t_b^3 \over g_b} &
\equiv t_{d5} \\
T_{KK6} R_1 =& {R_1 R_2^2 \over L_p^9} = {t_b^3 \over g_b^2} &
\equiv t_{s5} 
\end{array} \label{tdb} \ee
Note that under $s$-duality the $d$3 is inert and the $d$5 and $s$5 are
exchanged. 

We still have to find the limit in which the bulk physics decouples.
Keeping $t_b$ and $g_b$ finite, the Planck mass in 9 dimensions
is given by:
\[ M_p^7={R_1 R_2 \over L_p^9}=\left({t_b^2\over g_b}\right) {1 \over 
L_p^3}, \]
and goes to infinity when $L_p \rightarrow 0$. To keep the parameters
of $ii$b finite, we also have to take $R_1, R_2 \rightarrow 0$.

We now consider the NS5(A) approach. The parameters are the string length
$\tilde{l}_s$, the string coupling $\tilde{g}_A$ of type 
IIA theory and the radius $\tilde{R}_A$ of
the compact direction. The configurations, breaking 1/4 supersymmetry, 
leading to finite tension objects
in the world-brane of the NS5(A) are: F1$\cap$NS5(A), D2$\mapsto$NS5(A), 
D4$\mapsto$NS(A), D6$\mapsto$NS(5) and KK5(A)$\parallel$NS5(A).
In this framework the string tension $t_b$ is defined by the fundamental
string, namely $t_b=\tilde{l}_s^{-2}$. 
The little string coupling $g_b$ is found
by identifying the tension of the $d$1-brane from the configuration with 
the D2. We have: 
\be
t_{d1}=T_{D2}\tilde{R}_A={\tilde{R}_A \over \tilde{g}_A \tilde{l}_s^3} 
\equiv {t_b \over g_b}, \qquad \qquad
g_b={\tilde{g}_A \tilde{l}_s \over \tilde{R}_A}
\label{gnsb}
\ee
We obtain this picture from the previous one by dimensional reduction
on $R_1$, $R_1=\tilde{g}_A \tilde{l}_s$. 
The $\tilde{R}_A$ here is the previous $R_2$.
In this case the limit is taken performing $\tilde{g}_A \rightarrow 0$ and
$\tilde{R}_A \rightarrow 0$ at fixed $t_b$ and $g_b$. 
Note that the $s$-duality
in this picture is less straightforward to obtain from 10 dimensional
string dualities (one has to operate a TST duality chain).

Turning now to the KK5(B) picture, we find that, as in the type $ii$a
case, the little string theory is the reduction to the world-volume
of the KK5 of the physics of the objects that fit inside it. Thus
we simply identify $t_b$ with ${\tilde l}_s^{-2}$, $g_b$ with $g_B$, 
$s$-duality with S-duality, $f$1 with F1
and so on. As in the previous KK5 case, the limit in which the bulk
decouples involves taking the radius of the NUT direction to infinity.

We recapitulate the BPS spectrum of type $ii$b theory in the following table.
As for the $ii$a case, we list the different little branes and their mass 
considering now a compact
world-volume characterized by radii $\Sigma_i$ with $i=1 \dots 5$ and volume 
$\tilde{V}_5=\Sigma_1 \dots \Sigma_5$.
\[
\begin{array}{|c|c|} \hline 
\mbox{Brane} & \mbox{Mass} \\ \hline
w & {1 \over \Sigma_i} \\
& \\
f1 & \Sigma_i t_b \\
& \\
d1 & { \Sigma_i t_b \over g_b} \\
& \\
d3 & {\tilde{V}_5 t_a^2 \over \Sigma_i \Sigma_j g_b} \\
& \\
d5 & {\tilde{V}_5 t_a^3 \over  g_b} \\
& \\
s5 & {\tilde{V}_5 t_a^3 \over g_b^2} \\
& \\ \hline
\end{array} 
\]
\centerline{Table 3: mass of the BPS objects in $ii$b theory.}

We also summarize below the different ways to obtain $ii$b theory
and the relation between the parameters.
\[
\begin{array}{|c|c|c|c|} \hline & & & \\
& M5 & NS5(A) & KK5(B) \\  
& L_p, R_1, R_2 \rightarrow 0 & \tilde{g}_A, \tilde{R}_A \rightarrow 0 
& R_{NUT} \rightarrow \infty \\ & & & \\ \hline & & & \\
t_b & {R_1 \over L_p^3} & {1 \over \tilde{l}_s^2} & {1 \over {\tilde l}_s^2} 
\\ & & & \\
g_b & {R_1 \over R_2} & {\tilde{g}_A \tilde{l}_s \over \tilde{R}_A} & g_B 
\\ & & & \\
\hline \end{array} 
\]
\centerline{Table 4: definitions of $ii$b parameters.}

As most easily seen in the pictures using the NS5 or the KK5 branes,
there is a $t$-duality relating $ii$a and $ii$b little string theories.
It is simply the 10 dimensional T-duality between IIA and IIB, applied
on a direction longitudinal to the world-volume of the above-mentioned
branes. To be more specific, application of such a longitudinal T-duality
maps, say, the NS5(A) picture of $ii$b theory to the NS5(B) picture
of $ii$a theory, and similarly for the KK5 pictures.
The behaviour of the BPS objects is the same as in type II string theories:
KK momenta are exchanged with wound $f$1 strings, the $s$5 brane of one
theory is mapped to the one of the other theory, and
$dp$-branes become $d(p+1)$- or $d(p-1)$-branes for transverse or 
longitudinal $t$-dualities respectively. $ii$a and $ii$b theories
are thus equivalent when reduced to 5 space-time dimensions or less.

\subsection{m-theory}

As stated at the beginning of this section, there are two objects with
7 dimensional world-volume in M/type II theories: the D6-brane in type
IIA and the KK6 monopole in M-theory. The supersymmetry algebra is unique
and obviously non-chiral.

We first consider the D6 approach. Note that for the transverse space
to be asymptotically flat, we cannot have any compact transverse
dimension. The free parameters are thus the string length $l_s$ and
type IIA string coupling $g_A$. Already at this stage we know that
the theory on the world-volume will be characterized by only one
parameter (one is lost taking the appropriate limit which decouples the
bulk).

In this case, we have to consider configurations preserving $1/4$
supersymmetries with a brane within the D6-brane. The only branes of type
IIA for which this works are the D2- and the NS5-brane \cite{aeh}.
We identify them with the $m$2 and $m$5 branes. As it is necessary for
the definition of m-theory, only one parameter suffices to define 
both their tensions. Indeed we have:
\be
\begin{array}{c}
t_{m2}\equiv T_{D2} = {1\over g_A l_s^3}\equiv {1\over l_m^3} \\
\\
t_{m5}\equiv T_{NS5} = {1\over g_A^2 l_s^6} = {1\over l_m^6}
\end{array} \label{tm} \ee
$l_m$ is thus the characteristic length of m-theory, the analog of the 
Planck length in M-theory.

In order to decouple gravity, we send the 10 dimensional Planck mass
to infinity. Keeping $l_m$ finite, we have:
\[ M_p^8={1\over g_A^2 l_s^8} = {1 \over (l_m^6) l_s^2}, \]
and thus we have to take $l_s \rightarrow 0$ and $g_A \rightarrow \infty$.

In the KK6 approach, there are two configurations preserving $1/4$ of 
supersymmetry: M2$\subset$KK6 and M5$\subset$KK6. M2 and M5 are thus
respectively identified to $m2$ and $m5$, and $l_m=\tilde{L}_p$ where 
$\tilde{L}_p$ is
the eleven dimensional Planck length.
The KK6 monopole can be seen as the M-theoretic origin (and thus the
strong coupling limit) of the D6-brane. The radius of the NUT direction
is thus given by $R_{NUT}=g_A l_s=g_A^{2/3} \tilde{L}_p$. 
Therefore, the limit above
$g_A \rightarrow \infty$ becomes $R_{NUT} \rightarrow \infty$. Again, in this
limit the geometry becomes that of flat space.

It is interesting to note that here as in the former cases of $ii$a and
$ii$b theories, the KK monopole description is the more ``economic" one,
in the sense that one has to take only one limit. However, the other 
descriptions will be useful to make contact with Matrix theory
compactifications.

In the table below the masses of the different BPS objects of m-theory
are listed. Again we consider a compact volume $\tilde{V}_6=\Sigma_1 \dots 
\Sigma_6$. 
\[
\begin{array}{|c|c|} \hline 
\mbox{Brane} & \mbox{Mass} \\ \hline
w & {1 \over \Sigma_i} \\
& \\
m2 &{ \Sigma_i\Sigma_j \over l_m^3} \\
& \\
m5 & {\tilde{V}_6 \over \Sigma_i l_m^6} \\
& \\
\hline
\end{array} 
\]
\centerline{Table 5: mass of the BPS objects in m-theory.}

The different ways to obtain m-theory are shown below, along with
the relation between the parameters.
\[
\begin{array}{|c|c|c|} \hline & &  \\
& D6 & KK6  \\  
& {1\over g_A}, l_s \rightarrow 0 
& R_{NUT} \rightarrow \infty \\ & &  \\ \hline & & \\
l_m & g_A^{{1\over3}}l_s & \tilde{L}_p \\ & &  \\
\hline \end{array} 
\]
\centerline{Table 6: definitions of m-theory parameters.}

The duality between m-theory and $ii$a theory can now be made more 
precise. The relations between the parameters of m-theory 
compactified on the ``7th" direction of radius $R_c$ and $ii$a theory
are easily found comparing the tensions of the wrapped and unwrapped
$m$2 brane on one side, and of the $f1$ and $d2$ branes on the other
side. One finds no surprises:
\[ t_a={R_c \over l_m^3}, \qquad \qquad g_a=\left({R_c\over l_m}
\right)^{3/2}.\]
In the KK5(A) and KK6 picture, this is a direct consequence of the
relations between M and IIA theories. It is more amusing to see
that they indeed correspond to T-duality relations between IIA and IIB 
when one goes to the D5/D6 picture.

\subsection{Relation with Matrix theory compactification}

The little theories discussed above are relevant to the description 
of Matrix theory compactified on higher dimensional tori.

In the original conjecture \cite{bfss}, M-theory in the IMF is described
by the Matrix theory of a system of $N$ D0-branes, in the large $N$ limit.
The radius $R$ of the compact 11th direction which is used to go to the 
IMF and the 11 dimensional Planck length $l_p$ enter 
in the theory of D0-branes via the coupling and the string length
of the auxiliary IIA string theory to which the D0-branes belong.
If some of the remaining 9 space directions are compactified (on $T^d$ say),
one has to correctly include in the Matrix description
the additional BPS states that will fit into representations
of the U-duality group of compactified M-theory. A way to achieve this
is to take the system of D0-branes on $T^d$ and transform it into a
system of $N$  D$d$-branes completely wrapped on the dual
torus \cite{bfss,tay1}. 
Then naively one could
hope that all the physics of M-theory on $T^d$ would be captured
by the SYM theory in $d+1$ dimensions which is the low-energy effective
action of this system of D$d$-branes. For completeness we list here
the relations between quantities in the string theory in which the 
D$d$-branes live, and Matrix theory variables (see e.g. \cite{hana,eliz}):
\be
l_s^2={l_p^3\over R}, \quad \Sigma_i={l_p^3\over RL_i}, \quad
g_s={R^{3-d \over 2} l_p^{{3\over 2}(d-1)} \over V_d}, \quad
g_{YM}^2=g_s l_s^{d-3}={R^{3-d} l_p^{3(d-2)}\over V_d}. \label{genform}
\ee
$l_s$ and $g_s$ are respectively the string length and coupling;
$L_i$ and $\Sigma_i$ are the sizes of the torus in M-theory and in 
the auxiliary string theory picture respectively, and $V_d=L_1\dots L_d$; 
$g_{YM}^2$ is the SYM coupling, which is dimensionful in $d\neq 3$.
Note that in the end to make contact with M-theory on $T^d$ we have
to take the limits $R\rightarrow \infty$ and $L_i \rightarrow 0$ at fixed
$l_p$, together with the large $N$ limit.

Now for $d\geq 4$ the SYM is ill-defined because non-renormalizable,
and thus the SYM prescription for Matrix compactification seems
to break down. However, what we should consider as a model for the 
description of M-theory on a torus is really the ``theory on the 
D-brane" and not only its low-energy field theory limit. 
Furthermore, to be able
to consider a system of $N$ D$d$-branes on its own, one has to take a 
limit in which the bulk physics in the auxiliary string theory decouples.
This limit has to be compatible with the other limits discussed in the
paragraph above.

For Matrix theory on $T^4$, it turns out \cite{roza1}
that the theory of D4-branes
at strong string coupling coincides with a 6 dimensional (2,0)
supersymmetric field theory, which is the theory of $N$ M5-branes
in flat space \cite{stro}. For Matrix on $T^5$, the theory of D5-branes
at strong coupling is mapped \cite{sei2} by a IIB S-duality to the theory of
$N$ NS5-branes at weak coupling, which is a theory with string-like 
excitations. Finally, Matrix theory on $T^6$ is a theory of D6-branes
which, at strong coupling, becomes a theory of KK6-monopoles
\cite{brun,hana}. This 7-dimensional theory has membranes and, as we showed
above, has a well-defined structure which has been called m-theory.

We will show in the remainder of this section how all the ``phases" of
m-theory (i.e. its 7- and 6-dimensional versions) describe M-theory
on $T^6$, and how some particular limits of them yield back the
compactifications on $T^5$ and $T^4$. In other words, we find the
theories mentioned above \cite{roza1,sei2}
as limits of the $ii$a and $ii$b little string theories.

Specializing now to $d=6$, we consider first m-theory in the D6-brane
picture. We have for the string coupling:
\be g_A={l_p^{15/2}\over R^{3/2} V_6}. \label{gA} \ee
For the m-theory to be well-defined, its length scale $l_m$ has to be a fixed
parameter. Picking its value from Table 6, it takes the following 
expression in Matrix theory variables:
\be l_m^3={l_p^{12}\over R^3 V_6}. \label{lm} \ee
Note that the limits $V_6 \rightarrow 0$ and $R\rightarrow \infty$ have
to be taken simultaneously and in a definite way, in order to keep
a well-defined theory in this limit. Note also that m-theory is
valid only in the $g_A \rightarrow \infty$ limit, and this is compatible
with the above limits since we can re-express $g_A=(l_m/l_p)^3 R$.
Also $l_s^2=l_p^3/R \rightarrow 0$ as wanted.

Knowing \rref{lm} and the relations between $\Sigma$'s and $L$'s, we
can now translate the masses of the BPS states in m-theory into
masses of M-theory objects. We know in advance to which kind of objects
they will map to: since the BPS states break half of the supersymmetries
of the little theories, they correspond to objects of M-theory in the IMF
which break 1/4 of the supersymmetries. These are branes with travelling
waves in the 11th direction, i.e. longitudinal branes. The
remaining dimensions of these branes are wrapped on the $T^6$. One could 
also have deduced this from the fact that the energies of these states
will be proportional to $n$ the number of BPS little branes, and
independent of $N$. Since these objects are string-like in the 5 dimensional
supergravity to which M-theory is reduced, they should carry the 27
magnetic charges of this theory (i.e. they should fit into
the ${\bf \overline {27}}$ of the U-duality group $E_6(Z)$ \cite{huto}). 
We indeed find the following identifications: the 15 $m$2 wrapped membranes
are mapped to longitudinal M5-branes, the 6 momenta $w$ are mapped to
longitudinal M2-branes, and the 6 $m$5 states are longitudinally wrapped
KK6 monopoles (the NUT direction being always on the $T^6$). Their masses
can be easily computed from Table 5 and their Matrix couterparts
can be found in \cite{eliz}. All these 27 states can be found also
in the $ii$a and $ii$b pictures to be discussed below, although the
identification is less straightforward. This clearly convinces that 
the little string theories are 6-dimensional phases of a description
of M-theory on $T^6$.

We would also like to obtain the spectrum of the 27 electric charges in 5
dimensional supergravity (fitting into the {\bf 27} of $E_6(Z)$). These
correspond to completely wrapped branes in M-theory, or transverse branes
in the Matrix theory language (they can be represented as boosted branes). 
These objects preserve 16 supercharges in the Matrix model, 
and thus are totally supersymmetric states of the little theory. 
In the low-energy
SYM picture of the little theories, some of these transverse branes can
be associated to the electric and magnetic fluxes of the SYM \cite{tay2,gura}. 
However
the transverse M5-branes are missing from this description, which is
thus incomplete (note also that there are no BPS states in the SYM which
would represent the longitudinal KK6, or m-theory's $m$5). Going
back to the D6-brane picture, one can find all these half-supersymmetric
states by embedding in the D6-branes other branes of type IIA theory
in a way that they make a non-threshold bound state
(the archetype of these states is the supergravity solution of \cite{izq}).
These states can be found by chains of dualities from \cite{izq}
and are: F1$\subset$D6, D4$\subset$D6 and KK5$\subset$D6.
The energy of these states can also be found in \cite{eliz}. When there
are $N$ D6-branes and $n$ other branes inside them, this energy goes
like $n^2/N$. 

We now discuss the other pictures and the other little theories, along
with the relations between their parameters and the Matrix theory 
variables. It is clear that the parameters of the little theories,
once expressed in Matrix variables, will no longer depend on the
picture by which the little theory was defined. It will be however
interesting to check that the limit in which Matrix theory is a good
representation of M-theory coincides with the limit in the auxiliary
theory in which the
little theory is well-defined. As an example, the KK6 picture for m-theory
is related to the D6 picture by going from IIA to M on the NUT direction
of the KK6. Then if we call $\tilde{L}_p$ the Planck length of the auxiliary
M-theory (not to be confused with the M-theory that we are supposed
to describe, characterized by $l_p$), we have that $\tilde{L}_p=g_A^{1/3}l_s
\equiv l_m$ and $R_{NUT}=g_A l_s=(\tilde{L}_p/l_p)^3 R \rightarrow \infty$.

The $ii$a theory is most easily obtained going from the D6 to the D5
picture by T-duality. The reason to do this could be that one of the
radii of $T^6$ is much bigger than the others, and we might want
to decompactify it eventually. Then the parameters characterizing
the IIB auxiliary theory in which the D5 lives are given by:
\be g_B={l_p^6\over R V_5}, \qquad l_s^2={l_p^3\over R}, \qquad
R_B=L_6, \label{d5param} \ee
where $V_5=L_1\dots L_5$. The parameters of the little $ii$a string theory
can be easily extracted using Table 2:
\be g_a={V_5^{1/2}\over l_p^{3/2} L_6}, \qquad t_a={R^2 V_5\over l_p^9}.
\label{iiaparam} \ee
The limits of Matrix theory ($L_i \rightarrow 0$ and $R\rightarrow\infty$)
are compatible with keeping $g_a$ and $t_a$ finite. Note however
that if $L_6\rightarrow \infty$ instead, then $t_a$ remains fixed
while $g_a$ inevitably goes to zero. In this limit all the branes of $ii$a
except the $f$1 aquire an infinite tension and thus decouple. We are
left with a little string theory at zero coupling, which has exactly
the right number of states to describe Matrix theory on $T^5$. It has
indeed 5 winding plus 5 momentum BPS states, which together make up
the 10 longitudinal states of Matrix on $T^5$ \cite{sei2}.

To show that the $ii$ strings tend exactly to the description of
Matrix on $T^5$ given by Seiberg \cite{sei2}, we first go to
the NS5(B) picture of $ii$a strings. This is performed by an S-duality,
and we obtain for the IIB parameters:
\be \tilde{g}_B={1\over g_B}={R V_5\over l_p^6}, \quad
\tilde{l}_s^2=g_B l_s^2={l_p^9\over R^2 V_5}, \quad
\tilde{R}_B=R_B=L_6. \label{ns5bpar}\ee
We now see that $\tilde{g}_B=t_a l_p^3/R\rightarrow 0$ when $R\rightarrow
\infty$, and that this limit is independent of $L_6$. It thus comes out
of this picture that the little string theories proposed by Seiberg
to describe Matrix on $T^5$ are the zero coupling limit of the more
complete $ii$ little string theories that describe Matrix on $T^6$.

In order to go to the $ii$b theory, we perform a T-duality along, say,
the $\hat{5}$ direction. We obtain a NS5-brane in a IIA theory
characterized by:
\be
\tilde{g}_A={\tilde{g}_B \tilde{l}_s \over \Sigma_5}={RV_4^{1/2}L_5^{3/2}
\over l_p^{9/2}}, \quad  \tilde{l}_s^2={l_p^9\over R^2 V_5}, \quad
\tilde{R}_A=L_6, \label{ns5apar} \ee
with $V_4=L_1\dots L_4$. It is worth noting that from the $ii$b point
of view, the 5th direction has a radius:
\be \Sigma_5'={\tilde{l}_s^2\over \Sigma_5}={l_p^6\over R V_4}.
\label{5prime} \ee
This expression has forgotten all dependence on $L_5$, and thus we should
no longer think of the fifth direction of the NS5(A) brane
as related to the fifth direction of the original $T^6$. Moreover,
we can identify $\Sigma_5'=g_{YM(4+1)}^2$, as in \cite{roza1}.
The parameters of the $ii$b theory are given by:
\be
g_b={L_5\over L_6}, \qquad t_b={R^2 V_4 L_5\over l_p^9}. \label{iibparam}\ee
Of course, we could have computed this parameters without leaving
the little string theories, by $t$-duality from the $ii$a-theory.
For $L_6 \rightarrow \infty$ and $L_5 \rightarrow 0$, $t_b$ can be fixed
but $g_b \rightarrow 0$ and we recover the second string theory
with 16 supercharges proposed by Seiberg \cite{sei2}. 
It is worth noting that $\tilde{g}_A=t_b^{1/2}L_5$ and that the IIA
coupling vanishes in this case, but that in the opposite limit, which
is appropriate to compactification on $T^4$, we are at strong coupling.
We are thus led to consider the M5 picture of $ii$b strings.

The M5 picture is easily obtained by decompactification of a new
direction in the auxiliary M-theory, the radius of which we denote as $R_1$.
The parameters are thus:
\be R_1=\tilde{g}_A \tilde{l}_s=L_5, \quad R_2=L_6, \quad
L_p^3={l_p^9 \over R^2 V_4}. \label{mpar}\ee
If we want the bulk to decouple we have to impose $L_p\rightarrow 0$.
This combined with $t_b=L_5/L_p^3$ implies that $t_b$ is
finite, and we have a little string theory, only if $L_5, L_6 \rightarrow 0$.
If we want to recover Matrix theory on $T^4$, we have to take the opposite
limit. When $L_5, L_6 \rightarrow \infty$,
the tension of the little strings becomes very large, only the massless
modes contribute,
and we are left with a field theory of a special kind, which is however
still 6 dimensional. We have thus reproduced the results of 
\cite{roza1,roza2}.

As a last remark on this issue, note that we could have gone to the M5
picture from the D5 one through a T-duality on $\hat{5}$ which
would have transformed the D5 into a D4, and then elevating the latter
to an M5-brane. Though the labelling of the directions in the auxiliary
theory is clearly different in this M5 from the one of the previous
paragraph, when expressed in Matrix variables the quantities are exactly
the same. This is related to the fact shown in \rref{5prime}
that in the $ii$b picture
the ``base space" does not refer any more to the original $L_5$.

\section{Theories with 8 supercharges}

We propose in this section to define the little string theories with
(1,0) supersymmetry in 6 dimensions. Note that this is the highest
dimension in which a theory with 8 supercharges can live. 
We construct the (1,0) theories by analogy with the 10 dimensional
relation between $N=1$ and $N=2$ string theories.

In 10 dimensions, type I open string theory can be obtained from type
IIB string theory \cite{sagn,petr,polc}. 
One adds to the IIB theory an $\Omega 9$
orientifold yielding $SO$ open strings \cite{tasi}, 
and then adds 16 D9-branes to
have a vanishing net flux of D9 RR charge. This leads to an $N=1$
supersymmetric theory
with open strings carrying $SO(32)$ Chan-Paton factors. 
The two heterotic string theories are then obtained by dualities.
The $SO(32)$ heterotic theory is found by S-duality from the type I
(identifying the D1-brane in the latter to the fundamental heterotic
string of the former \cite{polc}). The $E_8 \times E_8$ heterotic
theory is obtained by T-duality from the $SO(32)$ one. 
The $E_8 \times E_8$ theory can also be derived from
M-theory compactified on $S^1/Z_2$ \cite{hora}.

Our strategy is the following: we define the theories with 8 supercharges
using the 5-branes of the $ii$ little string theories, and we then
show that the same pattern of dualities as in 10 dimensions arises.

Let us start with the $ii$b little string theory, where we can define 
a procedure very close to that of \cite{sagn,petr,polc}. 
In this theory we have
$d$5-branes (cfr. Table 3), which are Dirichlet branes for the little
$ii$b fundamental strings, filling the 6-dimensional
space-time. They are thus the analogue of the D9-branes of type IIB
theory. We now go to one of the precise pictures of section 2.2 to
analyze the structure of the theory defined by $ii$b in presence 
of a certain number $n$ of $d$5-branes.

If we take the KK5(B) picture (see Table 4), the $d$5-brane arises
from the $D=10$ D5-brane with its world-volume inside the KK5.
It is now straightforward to identify which BPS states of the $ii$b
theory survive the ``projection" due to the presence of the $d$5-branes.
From the 10-dimensional supersymmetry relations listed in the appendix,
we can see that only D1-branes can live at the same time within
the KK5 and the D5-branes. The closed $f$1, coinciding with the F1, is no
longer a BPS state, and the same occurs to the $d$3 and the $s$5. We
are thus left with a theory of open little strings (the open
IIB strings within the D5-brane), with a $d$1-brane BPS state. We 
propose to call this theory type $i$.

Note that along with the $n$ D5-branes, 
one can also add an $\Omega5$ orientifold plane\footnote{
Much in the same way as it was introduced in \cite{evan} in the 
context of brane configurations describing field theory dualities involving
$SO$ and $Sp$ groups.} 
without breaking further supersymmetry. Since there are still 3
non-compact transverse directions, the $SO$ or $Sp$ nature of the
orientifold and the number of D5-branes is not fixed by simple charge
flux arguments. Therefore, unlike the 10 dimensional case, here we can
have a priori arbitrary $U(n)$, $SO(2n)$ or $Sp(2n)$ gauge groups on 
the D5-branes.
The $\Omega 5$ defines an $\omega 5$ little orientifold plane for
the $ii$b theory. 

If there is only one KK5 brane, the gauge group discussed above
corresponds to the gauge group of the little type $i$ string theory.
On the other hand, if there are $N$ coinciding KK5 branes (as it
should be in a Matrix theory perspective), this issue is more subtle.
We return on this at the end of the section. 

In order to define a (1,0) closed string theory, we can simply apply
the $s$-duality of section 2.2 to the type $i$ theory. This
duality maps the $d$1 branes to the $f$1 little strings, and most
notably the $d$5-branes to the $s$5-branes. The only BPS states
of this theory are thus the $f$1. We call this theory $h_b$. We could
have directly found this $h_b$ theory from the $ii$b one
by piling up $n$ $s$5-branes. If we are allowed to define the $s$-dual
of the $\omega5$ orientifold, then this procedure is reminiscent
of the one used by Hull \cite{hull} to obtain the heterotic $SO(32)$ theory
from type IIB. The possible gauge groups of the $h_b$ theory are
the same as the ones for type $i$ theory.

There is still a 5-dimensional object in the little string theories
that could be used to define a new (1,0) theory, namely the $s$5-brane
of the type $ii$a theory. 
Taking the KK5(A) picture, we obtain this theory piling up $n$
NS5(A)-branes inside its world-volume. However in this case the 
gauge symmetry, even in the simplest case of a single KK5, is
unclear. This is related to the present lack of a definition of a gauge
theory associated to the (2,0) theory of $n$ NS5(A)- or M5-branes.
We call this little string theory $h_a$. It is $t$-dual to the
$h_b$ one. 

Elevating the picture of a KK5 parallel to NS5-branes in type 
II theory to M-theory, we find a KK6 with M5 branes defining a domain wall,
or boundary of its 7-dimensional world-volume. This is m-theory
with $m$5-branes. Thus the $h_a$ theory can be seen as an m-theory
compactification in presence of $m$5-branes.
This description is very rough and schematic, 
but could be related
to a 7-dimensional analogue of the Horava-Witten mechanism \cite{hora}
to obtain the $E_8 \times E_8$ heterotic string theory
(although in \cite{hora} the 9-dimensional objects are 
really boundaries rather than branes). 

We thus see that the pattern of dualities that arises between the
theories with 8 and 16 supercharges is very similar to the one
between $N=1$ and $N=2$ string theories in 10 dimensions. We list in the 
table below the main characteristics of the (1,0) little string theories.
\[
\begin{array}{|c|c|c|} \hline 
\mbox{Theory} & \mbox{Defined by:} & \mbox{BPS objects}   \\ \hline
i  & ii\mbox{b} + d5 & d1 \\
h_a  & ii\mbox{a} + s5 & f1 \\
h_b  & ii\mbox{b} + s5 & f1 \\
\hline
\end{array} 
\]
\centerline{Table 7: main characteristics of the theories with 8 
supercharges}

We now turn to the discussion of some speculative points related
to the theories discussed above.

Consider first the case where the little theories are defined by $N$ branes
of the same kind. For simplicity, we specialize to the KK5 picture. 
In that case, our approach does not help in 
clarifying which gauge group characterizes the little (1,0) theory.
The answer to this problem is however likely to be non-trivial. This can
be seen as follows. Take for instance the type $i$ theory. The configuration
discussed above to define it involved $N$ KK5 branes parallel to $n$
D5-branes. After a T-duality on the NUT direction we end up with
$N$ NS5-branes within $n$ D6-branes. This is related by T-dualities
to the configuration studied by Hanany and Witten \cite{hawi} of
D3-branes suspended between NS5-branes. In our case, the direction
of the D-branes perpendicular to the NS5-branes is compact
(as considered in e.g. \cite{hori,witm}). If the NS5 branes were distributed 
along this compact direction instead of being coincident, the gauge 
group would have been $U(n)^N$ \cite{hori}. In the limit in which
the NS5-branes are 
taken to coincide, it is not clear what gauge theory we get.

The $h_a$ and $h_b$ theories
can also be defined by $N$ NS5-branes, with the $s$5-branes provided 
by $n$ KK5 monopoles (this is obtained by a T-duality from the KK5
picture considered before; here the NS5 and the KK5 branes play the
opposite r\^ole). Since a background of multiple KK5 branes
can be related to an ALE space \cite{sen}, the $h_a$ and $h_b$ theories
should be connected to those studied in \cite{intri}.

Seiberg \cite{sei2} defines (1,0) little string theories from the
world-volume of the 5-branes in the two heterotic string
theories. These little theories have however a global $SO(32)$ or
$E_8 \times E_8$ symmetry, which is unlikely to arise in our
cases. The (1,0) theories of \cite{sei2} seem thus different from those
discussed in this section (in the sense that it should not be possible
to derive them from a pure type $ii$ little string framework).

It would be interesting to have an interpretation of the $i$, $h_a$ and
$h_b$ theories in terms of compactifications of Matrix theory on
6 dimensional manifolds breaking half of the supersymmetries.

As a side remark, it is worth noting that 
the 5-branes of the little theories play a crucial r\^ole in the 
interplay between theories with 16 and 8 supercharges. 
By analogy, 9-branes in M-theory and in type
II theories might be interesting to study.  
The existence of an M9-brane and NS-like 9-branes of type IIA and IIB theory
was indeed discussed in \cite{hull}.

\section{Discussion}

We have given in this paper a description of little theories in 6 and
7 dimensions. Our analysis is entirely
based on the spectrum of BPS states present in each one of these
theories. The focus on BPS states is partly motivated by the 
application of these little theories to the Matrix theory description
of M-theory compactifications, and to the necessity to recover
the right U-duality group. If we want to understand more deeply the nature
of M-theory, a study of these non-critical
string theories and m-theory beyond the BPS analysis is certainly mandatory. 
A promising avenue is to
consider a Matrix approach to these theories, as it was initiated recently in
\cite{aha1,hana,sethi,aha2,ganor,sethi2} for several related theories.

A full quantum and possibly non-perturbative formulation of these theories will
elucidate the relation between the little string theories or m-theory and
their low-energy effective action, which must not contain gravity.
In other words, this formulation should reproduce the low-energy effective
action of the branes used to define the little theories. It may also help
in understanding the full structure of the (2,0) field theory in 6 dimensions.
An interesting remark is that if we consider the (2,0) and the (1,1) 
six dimensional field theories as the low-energy effective actions
of type $ii$b and $ii$a string theories, then we can observe that
both are independent of the (little) string couplings $g_b$ and $g_a$.
This is because the first has no coupling at all, and the second has
a SYM coupling $g_{YM}^2=t_a^{-1}$. This is another characteristic of
the 6 dimensional strings which differentiates them from their 10 dimensional
sisters.

\subsection*{Acknowledgements}

We warmly thank Fran\c{c}ois Englert for numerous interesting discussions.
We also benefited from conversations with M.~Henneaux, A.~Sevrin and 
R.~Siebelink.

\appendix

\section{Supersymmetry properties and tensions of \break \hfill branes in 10 
and 11 dimensions}

In this appendix, we give a list of the projections 
imposed on the supersymmetric 
parameters of the theory when there is a brane in the background. We also
give the tensions of the branes.

In M-theory, we have one Majorana supersymmetric parameter $\epsilon$. The 
$\Gamma_M$ matrices are such that the one corresponding to the 11th direction
satisfies $\Gamma_{11}=\Gamma_0 \dots \Gamma_9$. We have the following
relations (the numbers between brackets indicate the directions longitudinal 
to
the brane, and W[1] stands for a travelling wave or KK momentum in the 
direction $\hat 1$):
\begin{eqnarray}
\mbox{W[1]:} & \qquad & \epsilon=\Gamma_0 \Gamma_1 \epsilon \nonumber\\
\mbox{M2[1,2]:} & \qquad & \epsilon=\Gamma_0 \Gamma_1 \Gamma_2 \epsilon 
\nonumber\\
\mbox{M5[1..5]:} & \qquad & \epsilon=\Gamma_0 \dots \Gamma_5 \epsilon 
\label{msusy}\\
\mbox{KK6[1..6]:} & \qquad & \epsilon=\Gamma_0 \dots \Gamma_6 \epsilon 
\nonumber\\
\mbox{M9[1..9]:} & \qquad & \epsilon=\Gamma_0 \dots \Gamma_9 \epsilon 
\nonumber
\end{eqnarray}
Note that there are no other combinations of the $\Gamma_M$ matrices which
square to the identity. These relations can be obtained from the 
11 dimensional supersymmetry algebra including tensorial central charges
\cite{demo,hull}.
Discarding all numerical factors, the tensions of these objects are given as 
follows.
The quantum of mass of a KK momentum on a compact direction of radius $R$ is:
\be M_{W}={1 \over R}. \label{mwave} \ee
If $L_p$ is the 11 dimensional Planck length, the tensions of the M2 and M5
branes are:
\be T_{M2}= {1 \over L_p^3}, \qquad T_{M5}={1 \over L_p^6}. \label{tmbranes}
\ee
The tension of a KK6 monopole with a transverse NUT direction of radius 
$R_N$ is:
\be T_{KK6}={R_N^2 \over L_p^9}. \label{tkk6} \ee
This can be easily obtained from the tension of a D6-brane. We do not 
discuss here the tension of the M9, which is not used in this paper.

In type II theories, there are 2 Majorana-Weyl spinors $\epsilon_L$ and
$\epsilon_R$ (with reference to the string origin of these susy generators). 
They satisfy the chirality conditions:
\[ \epsilon_L=\Gamma_{11} \epsilon_L, \qquad \epsilon_R=\eta \Gamma_{11} 
\epsilon_R, \]
with $\eta=+1$ for IIB theory and $\eta=-1$ for IIA theory. The supersymmetry
projections are the following (we denote by F1 the fundamental strings of each
theory):
\begin{eqnarray}
\mbox{F1[1]:} & \qquad & \left\{ \begin{array}{rcl} \epsilon_L&=&\Gamma_0
\Gamma_1 \epsilon_L \\ \epsilon_R&=&-\Gamma_0 \Gamma_1 \epsilon_R 
\end{array} \right. \nonumber \\
\mbox{W[1]:} & \qquad & \left\{ \begin{array}{rcl} \epsilon_L&=&\Gamma_0
\Gamma_1 \epsilon_L \\ \epsilon_R&=&\Gamma_0 \Gamma_1 \epsilon_R
\end{array} \right. \nonumber \\
\mbox{NS5[1..5]:} & \qquad & \left\{ \begin{array}{rcl} \epsilon_L&=&\Gamma_0 
\dots
\Gamma_5 \epsilon_L \\ \epsilon_R&=&-\eta \Gamma_0 \dots \Gamma_5 \epsilon_R
\end{array} \right. \label{iisusy} \\
\mbox{KK5[1..5]:} & \qquad & \left\{ \begin{array}{rcl} \epsilon_L&=&\Gamma_0 
\dots
\Gamma_5 \epsilon_L \\ \epsilon_R&=&\eta \Gamma_0 \dots \Gamma_5 \epsilon_R
\end{array} \right. \nonumber \\
\mbox{D$p$[1..$p$]:} & \qquad & \epsilon_L=\Gamma_0 \dots \Gamma_p \epsilon_R
\nonumber
\end{eqnarray}
Note that the relations for IIA theory are obtained from those of M-theory
compactifying on the 11th direction. $\Gamma_{11}$ plays thus the r\^ole of
the chiral projector in 10 dimensions, and the supersymmetry parameters are
related by $\epsilon_{L(R)}={1 \over 2}(1\pm\Gamma_{11})\epsilon$.
Also the relations of IIA and IIB theories are related by T-duality,
namely under a T-duality over the $\hat{\i}$ direction the susy parameters 
transform (see e.g. \cite{tasi})
as $\epsilon_L \rightarrow \epsilon_L$ and $\epsilon_R \rightarrow
\Gamma_i \epsilon_R$.

The mass of a KK mode W is as in \rref{mwave}. Type II string theories are
both characterized by the string length $l_s=\sqrt{\alpha'}$ and by the string
coupling constant $g$. The tension of the fundamental string is:
\be T_{F1}={1 \over l_s^2}. \label{tfund} \ee
The tension of the solitonic NS5 branes is given by:
\be T_{NS5}={1 \over g^2 l_s^6}. \label{tns} \ee
The KK5 monopole has a tension of:
\be T_{KK5}={R_N^2 \over g^2 l_s^8}, \label{tkk5} \ee
where $R_N$ is the radius of the NUT direction. Finally the tensions of the
D$p$-branes are given by:
\be T_{Dp}={1 \over g l_s^{p+1}}. \label{tdbrane} \ee

\end{document}